\documentstyle[epsf]{article}

\font\tenrm=cmr10
\font\tenit=cmti10
\font\elevenbf=cmbx10 scaled\magstep 1
\font\elevenrm=cmr10 scaled\magstep 1
\font\elevenit=cmti10 scaled\magstep 1

\renewenvironment{thebibliography}[1]
 { \elevenrm
   \begin{list}{\arabic{enumi}.}
    {\usecounter{enumi} \setlength{\parsep}{0pt}
     \setlength{\itemsep}{3pt} \settowidth{\labelwidth}{#1.}
     \sloppy
    }}{\end{list}}

\begin{document}
\font\fortssbx=cmssbx10 scaled \magstep2
\hbox to \hsize{
%\special{psfile=/NextLibrary/TeX/tex/inputs/uwlogo.ps
%			      hscale=8000 vscale=8000
%			       hoffset=-12 voffset=-2}
\hskip.5in \raise.1in\hbox{\fortssbx University of Wisconsin - Madison}
\hfill\vbox{\hbox{\bf MAD/PH/798}
            \hbox{October 1993}} }

\begin{center}
\vglue 0.6cm
{
 {\elevenbf        \vglue 10pt
               YUKAWA COUPLING EVOLUTION IN SUSY GUTS\footnote{Talk presented
by MSB at the
XVI Kazimierz Meeting on Elementary Particle Physics, 24-28 May 1993,
Kazimierz, Poland.
}%\\
%               \vglue 3pt
\\}
\vglue 1.0cm
{\tenrm V.~Barger, M.~S.~Berger,
and P.~Ohmann \\}
\baselineskip=13pt
{\tenit Physics Department, University of Wisconsin\\}
\baselineskip=12pt
{\tenit Madison, WI 53706, USA\\}}

\vglue 0.8cm
{\tenrm ABSTRACT}

\end{center}

\vglue 0.3cm
{\rightskip=3pc
 \leftskip=3pc
 \tenrm\baselineskip=12pt
 \noindent
The scaling behavior and fixed points in the evolution of fermion Yukawa
couplings and mixing angles are discussed. The relevance of fixed points in
determining the top quark mass is described.}

\vglue 0.4in
\baselineskip=14pt
\elevenrm
The unification of gauge couplings is a well-known feature of unified theories.
To understand the masses of the observed fermions requires ideas beyond the
standard model. With certain assumptions about
the interactions in the GUT theory,
relationships arise between various fermion masses. A mild assumption
of minimality yields the well-known relation$^1$ between the bottom and tau
Yukawa couplings at the GUT scale
$\lambda _b=\lambda _\tau $. A more ambitious program is the attempt to
formulate ansatze for the Yukawa coupling matrices, and obtain relationships
between fermion masses and CKM relations as well.

Unlike the gauge couplings or the CKM matrix elements, infrared
fixed point solutions$^{2-7}$
exist for the Yukawa coupling. Since the top quark is the only known fermion
with a mass of order of the electroweak scale, it is usually the only
particle for which the fixed point solution is relevant. The bottom and tau
Yukawa couplings can be large in a multi-Higgs doublet model, where the
smallness of the masses can be attributed to the properties of the vacuum.

The fixed point of the top quark Yukawa coupling in the minimal supersymmetric
standard model (MSSM)
can be obtained approximately
by
setting
\begin{eqnarray}
{{d\lambda _t}\over {dt}}&=&{{\lambda _t}\over {16\pi ^2}}
\Bigg (-\sum c_ig_i^2+6\lambda _t^2+\lambda _b^2
\Bigg )=0\;,
\end{eqnarray}
with $c_1=13/15$, $c_2=3$, $c_3=16/3$.
This is only accurate to about 10\% in practice
because the gauge couplings are themselves evolving.
An analysis of the two-loop RGEs in the MSSM using experimental
input for the gauge couplings yields an {\it effective} fixed point of
$\lambda _t^{fp}\simeq 1.1$ near the electroweak scale $\mu =M_Z$ as shown in
Figure 1. Top quark Yukawa couplings exceeding the fixed point value at the
GUT scale evolve rapidly to the fixed point, while the approach from below
is more gradual.

\begin{center}
\epsfxsize=2.8in
\hspace*{0in}
\epsffile{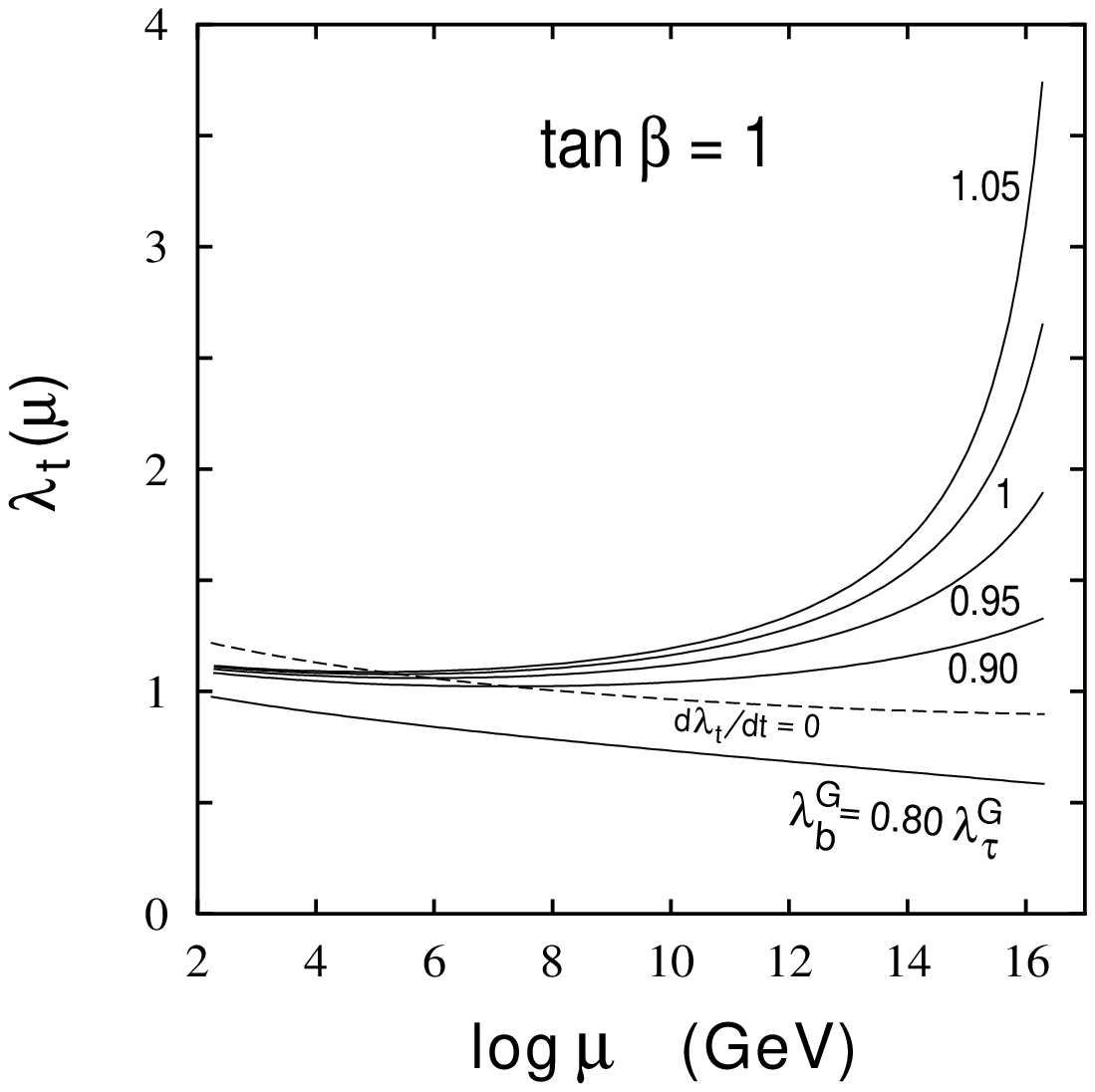}

\parbox{5.5in}{\small Fig.~1. The top quark Yukawa coupling evolves
rapidly to the
effective fixed point value from above. The constraint $d\lambda _t/dt=0$
varies with scale because the gauge couplings are evolving.}
\end{center}

The prediction for the $m_b/m_{\tau }$
ratio provides motivation for the fixed point solution.
If we take as inputs $m_{\tau }=1.784$ GeV and the running mass
$m_b(m_b)=4.25$ GeV, a large top quark Yukawa coupling is needed to
counteract the evolution due to the gauge couplings which alone yield too large
a value for the $m_b/m_{\tau }$ ratio.
In the standard model the effective fixed point solution
implies that the top quark is heavy $m_t > 200$ GeV. In the MSSM the
Yukawa coupling must be large ($\simeq 1$) at the electroweak scale, implying
a linear correlation between $m_t$ and $\sin \beta $ (neglecting contributions
from $\lambda _b$ and $\lambda _{\tau}$ which have a significant effect only
for very large $\tan \beta$),
\begin{eqnarray}
m_t(m_t)&=&(192 {\rm GeV})\sin \beta \;,
\label{mt}
\end{eqnarray}
where $\tan \beta $ is the ratio of the vevs of the two
Higgs doublets in the MSSM. This solution is indicated as the bands shown in
Figure 2.
As $\alpha _3(\mu )$ is increased, $\lambda _t(\mu )$ must be correspondingly
increased to preserve the $m_b/m_{\tau}$ prediction. Hence for larger
input $\alpha _3(M_Z)$, the solutions tend to display more strongly the fixed
point character as shown in Figure 2.
The fixed point does {\it not} require that $\tan \beta $ be small,
but allows for large $\tan \beta $ if $m_t$ is sufficiently large.
However if  $m_t^{\rm pole}$ is below $160$ GeV, the fixed point gives
$\tan \beta < 2$ with interesting consequences for Higgs boson
phenomenology$^{6}$.

Corrections to the Yukawa couplings at the GUT scale arise similarly to the
much-discussed threshold corrections to the gauge couplings. These two
types of threshold corrections are correlated since they
arise from the spectrum of massive states in the GUT theory$^{7}$.

The fixed point solution is largely independent of the
supersymmetric spectrum and therefore of the exact nature of
supersymmetry breaking. It is known that the minimal supergravity models
require that $\tan \beta >1$ in order to achieve the required symmetry
breaking. This is manifested in the (tree-level) relation
\begin{eqnarray}
{1\over 2}M_Z^2&=&{{m_{H_1}^2-m_{H_2}^2\tan ^2\beta }
\over {\tan ^2\beta -1}}-\mu ^2 \;. \label{treemin1}
\end{eqnarray}
Here $m_{H_1}$ and
$m_{H_2}$ are the masses of the two Higgs doublets, and $\mu $ is the Higgs
mass parameter appearing in the superpotential.
For $\tan \beta $ near one, $\mu $ must be large to permit the
substantial cancellation to achieve the correct $M_Z^{}$.
Various criteria can be chosen to determine where this cancellation becomes
unnnatural; this choice is largely a matter of taste.
This unnaturalness is ameliorated somewhat with the inclusion of one-loop
corrections$^{8-11}$. In any case, the large value of $\mu $ has
interesting consequences for the supersymmetric spectrum.
The $\tan \beta = 1$ direction in field space is a D-flat direction of the
supersymmetric theory. Consequently an associated Higgs eigenstate
is precisely massless at tree level for $\tan \beta = 1$.

\begin{center}
\epsfxsize=5.875in
\hspace*{0in}
\epsffile{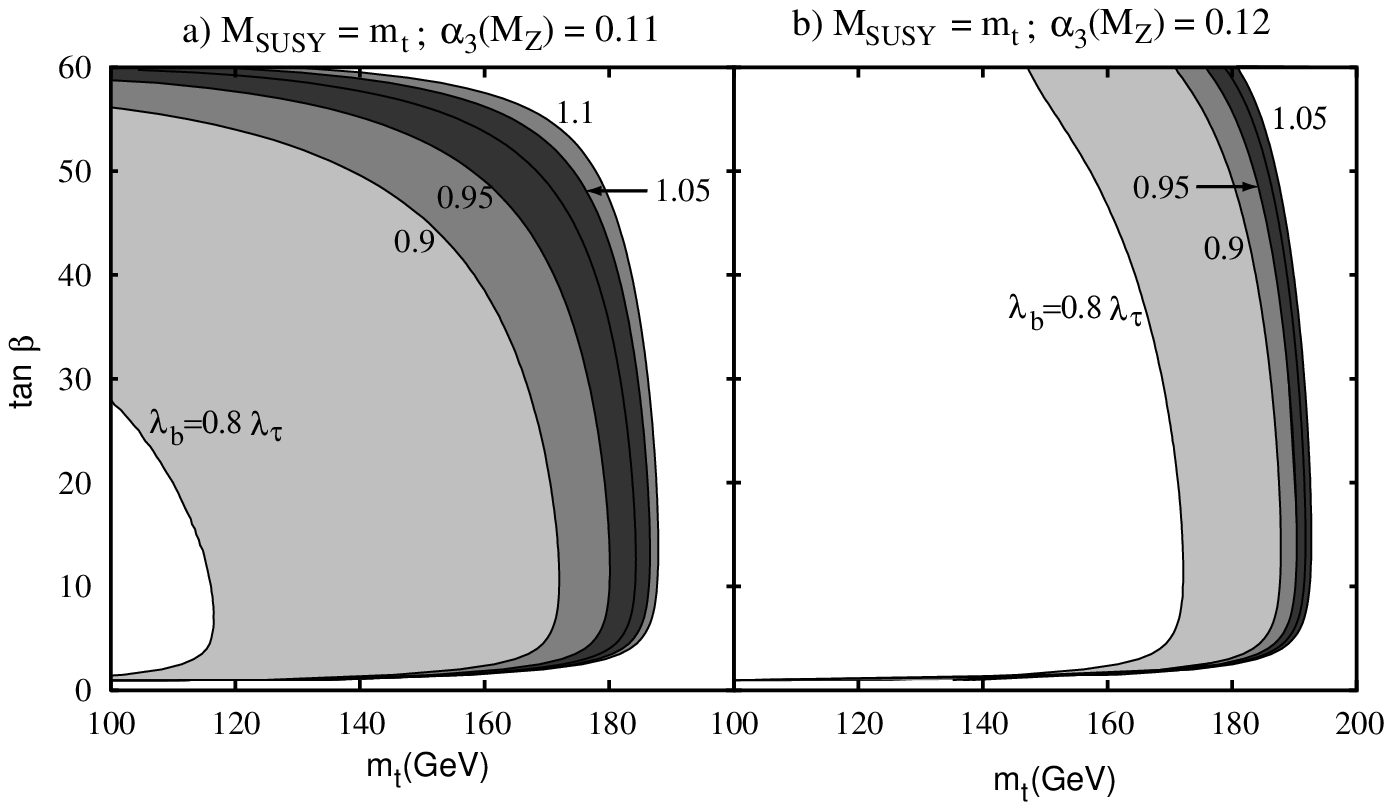}

\parbox{5.5in}{\small Fig.~2. The effect of threshold corrections on the
Yukawa coupling unification condition $\lambda _b(M_G^{})=\lambda _{\tau }$
 with
$m_b(m_b)=4.25$ GeV for $\alpha _3(M_Z^{})=0.11$ and $0.12$. }
\end{center}

When there is a hierarchy of masses in the Yukawa matrices, the evolution of
the quark masses and CKM mixing angles is given as a simple scaling.
If only one generation is heavy (with Yukawa couplings $\simeq 1$) and the
mixing angles of this generation with the light generations is small, the
evolution is given by scaling equations.
The CKM mixing angles evolve as$^{12}$
\begin{equation}
{{dW_1}\over {dt}}=-{{W_1}\over {8\pi ^2}}
\left (\lambda _t^2
+\lambda _b^2\right ) \;, \label{dW1dt}
\end{equation}
in the MSSM where $W_1=|V_{cb}|^2, |V_{ub}|^2, |V_{ts}|^2, |V_{td}|^2$, the
CP-violation parameter $J$
and
\begin{equation}
{{dW_2}\over {dt}}=0\;, \label{dW2dt}
\end{equation}
where $W_2=|V_{us}|^2, |V_{cd}|^2, |V_{tb}|^2, |V_{cs}|^2, |V_{ud}|^2$.
The solution of Eq.~(\ref{dW1dt}) is given by the scaling equation
\begin{equation}
W_1(M_G^{})=W_1(\mu )\exp \left \{ -{1\over {8\pi ^2}}\int\limits_{\mu}^{M_G}
\left (\lambda _t^2
+\lambda _b^2\right )\;
d\ln\mu' \right \}
\label{scal} \;.
\end{equation}
The lightest two generations do not affect the evolution, and
one does not need the mixing between the first two generations to be small
for the universal scaling described above to occur.
This makes the scaling universality an especially good approximation
since the Cabbibo angle is the largest of the quark mixings.
The scaling behavior can be demonstrated to all orders in perturbation
theory.

\vglue 0.6cm
{\elevenbf\noindent Acknowledgements}
\vglue 0.2cm
This research was supported
in part by the University of Wisconsin Research Committee with funds granted by
the Wisconsin Alumni Research Foundation, in part by the U.S.~Department of
Energy under contract no.~DE-AC02-76ER00881, and in part by the Texas National
Laboratory Research Commission under grant nos.~RGFY9273 and FCFY9302.
PO was supported in part by an NSF Graduate Fellowship.

\vglue 0.6cm
{\elevenbf\noindent References}
\vglue 0.2cm


\begin{thebibliography}{9}
\bibitem{CEG} M.~Chanowitz, J.~Ellis and M.~Gaillard,
   {\elevenit Nucl. Phys.} {\bf B128} (1977) 506.

\bibitem{Pendleton} B.~Pendleton and G.~G.~Ross,
   {\elevenit Phys. Lett.} {\bf B98}  (1981) 291; C.~T.~Hill,
   {\elevenit Phys. Rev.} {\bf D42} (1981) 691.

\bibitem{Knowles} C.~D.~Froggatt, I.~G.~Knowles and R.~G.~Moorhouse,
   {\elevenit Phys. Lett.} {\bf B249} (1990) 273;
   {\elevenit Phys. Lett.} {\bf B298} (1993) 356.

\bibitem{BBO1} V.~Barger, M.~S.~Berger, and P.~Ohmann, {\elevenit Phys. Rev.}
   {\bf D47} (1993) 1093; V.~Barger, M.~S.~Berger, T.~Han and M.~Zralek,
   {\elevenit Phys. Rev. Lett.} {\bf 68} (1992) 3394.

\bibitem{BCPW} M.~Carena, S.~Pokorski, and C.~E.~M.~Wagner,
{\elevenit Nucl. Phys.} {\bf B406} (1993) 59; W.~Bardeen, M.~Carena,
S.~Pokorski, and C.~E.~M.~Wagner, Munich preprint MPI-Ph/93-58.

\bibitem{bbop} V.~Barger, M.~S.~Berger, P.~Ohmann, and R.~J.~N.~Phillips,
   {\elevenit Phys. Lett.} {\bf B314} (1993) 351.

\bibitem{LP} P.~Langacker and N.~Polonsky, University of Pennsylvania preprint
UPR-0556-T (1993).

\bibitem{aspects} S.~Kelley, J.~L.~Lopez, D.~V.~Nanopoulos, H.~Pois, and
K.~Yuan, {\elevenit Nucl. Phys.} {\bf B398} (1993) 3.

\bibitem{op} M.~Olechowski and S.~Pokorski, {\elevenit Nucl. Phys.} {\bf B404}
(1993) 590.

\bibitem{dcc} B.~de~Carlos and J.~A.~Casas, {\elevenit Phys. Lett.}
{\bf B309} (1993)
320.

\bibitem{bbo3} V.~Barger, M.~S.~Berger, and P.~Ohmann, in preparation.

\bibitem{ckmrun} E.~Ma and S.~Pakvasa, {\elevenit Phys. Lett.}
{\bf B86}  (1979) 43,
{\elevenit Phys. Rev.} {\bf D20} (1979) 2899;
K.~Sasaki, {\elevenit Z. Phys.} {\bf C32}
(1986) 149;
K.~S.~Babu, {\elevenit Z. Phys.} {\bf C35} (1987) 69;
M.~Olechowski and S.~Pokorski, {\elevenit Phys. Lett.}
{\bf B257} (1991) 388;
V.~Barger, M.~S.~Berger, and P.~Ohmann, {\elevenit Phys. Rev.}
   {\bf D47} (1993) 2038;
K.~S.~Babu and Q.~Shafi, {\elevenit Phys. Rev.} {\bf D47} (1993) 5004.

\end{thebibliography}
\end{document}